\documentclass[aps,prl,floatfix,twocolumn,showpacs,reprint,superscriptaddress]{revtex4-1}
\usepackage{amsfonts}
\usepackage{mathrsfs}
\usepackage{amsmath}
\usepackage{color}
\usepackage{graphicx}
\usepackage{bm}
\usepackage{amssymb}
\usepackage{xspace}
\usepackage{epstopdf}
\usepackage{dcolumn}
\usepackage{longtable}
\usepackage{multirow}
\usepackage[colorlinks=true, letterpaper=true, pdfstartview=FitV, linkcolor=blue, citecolor=blue, urlcolor=blue]{hyperref}
\begin{document}

\preprint{APS/123-QED}

\title{Magnetic Response Functions in Landau Levels}

\author{Yang Gao}
\affiliation{Department of Physics, The University of Texas at Austin, Austin, Texas 78712, USA}

\author{Qian Niu}
\affiliation{Department of Physics, The University of Texas at Austin, Austin, Texas 78712, USA}
\affiliation{International Center for Quantum Materials, Peking University, Beijing 100871, China}

\date{\today}

\begin{abstract}
We propose a new quantization rule which generates Landau levels consistent with the zero-field magnetic response functions from the semiclassical theory. It reproduces the Onsager's rule in the leading order, and re-formulates corrections to the Onsager's rule from the Berry phase and magnetic moment effect in terms of one single  magnetic response: the zero-field magnetization. It can yield higher order corrections by including successively magnetic susceptibility and higher order magnetic response functions. In application, it can be easily applied to obtain Landau levels in lattice models. Moreover, it provides an experimental method of measuring different magnetic response functions directly from the measurement of Landau level fan diagram or Hofstadter spectrum.
\end{abstract}


\maketitle

Quantization of electronic states into Landau levels by a magnetic field
gives rise to oscillations in magnetization (de Haas-van Alphen effect)
and conductivity (Shubnikov-de Haas effect), which contains a wealth of information 
about the band structure as well as geometric properties of the Bloch states.  
The shape of Fermi surfaces \cite{Mermin1976} of metals and semiconductors were obtained in this 
way based on Onsager's quantization rule \cite{Onsager1952}.  Then, a shift of the Landau level index to leading order in the magnetic field has been studied from very early on \cite{Roth1966}. Later, this zeroth order correction is related to the Berry phase \cite{Wilkinson1984,Rammal1985,Mikitik1999} and the magnetic moment \cite{Chang1996,Xiao2010, Carmier2008, Fuchs2010} of the Bloch states.  Further corrections from the magnetic field remains rather obscure in physical significance and difficult in numerical calculations.

It is well known that from the free energy of the Landau level spectrum, various zero-field magnetic response functions such as magnetization and magnetic susceptibility can be derived, and they are important intrinsic properties of solids, as illustrated in the study of topological materials such as graphene \cite{Koshino2007, Schober2012, Santos2011, Yang2015,Raoux2015} and Dirac semimetals \cite{Koshino2010}. However, the following inverse questions have rarely been discussed: if the zero-field magnetic response functions are given, can we find a consistent Landau level spectrum, and if we can, how accurate is this spectrum?

\begin{figure}
\setlength{\abovecaptionskip}{0pt}
\setlength{\belowcaptionskip}{0pt}
\scalebox{0.32}{\includegraphics*{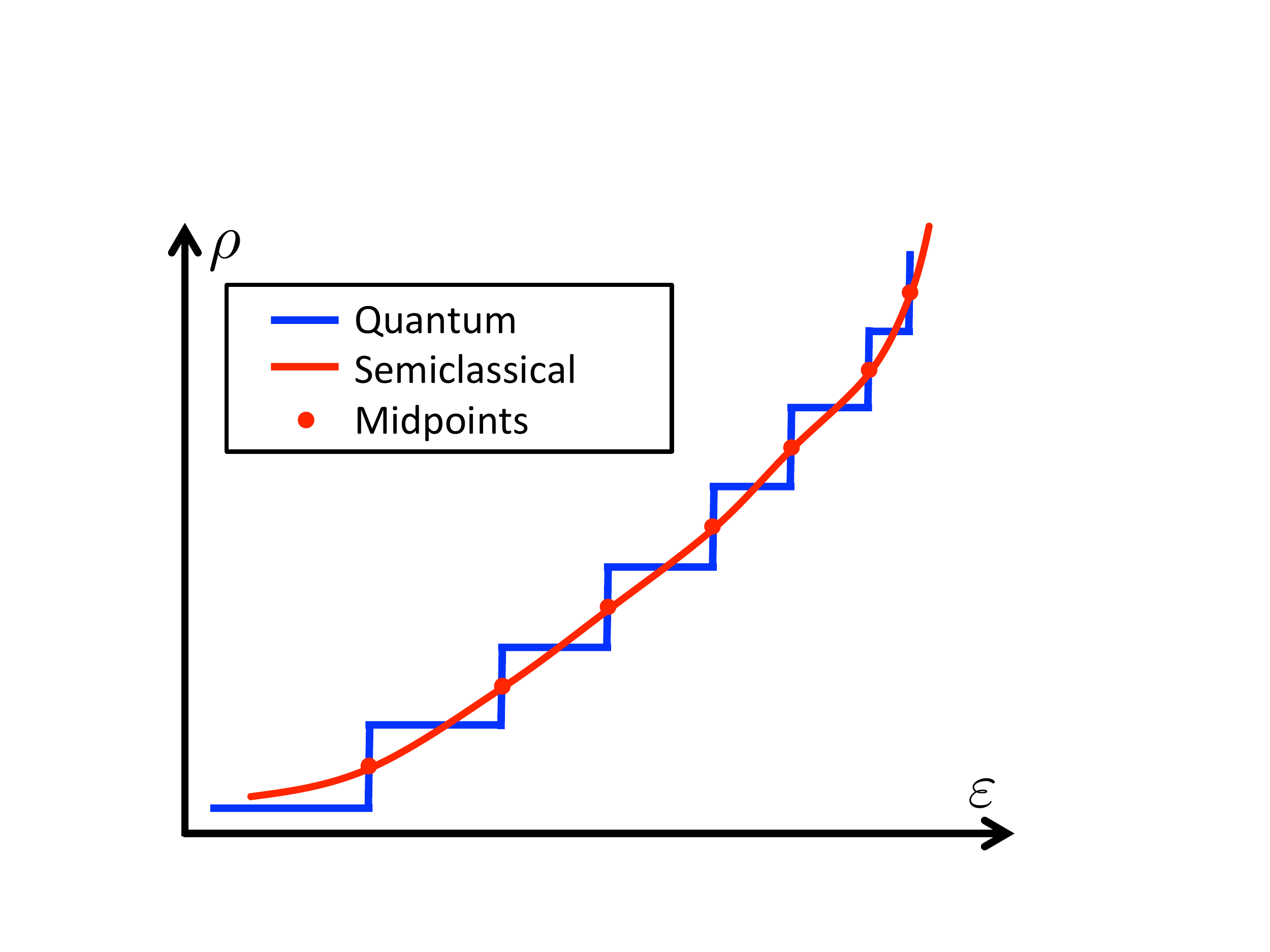}}
\caption{(color online) The intersection theorem.}
\label{fig_int}
\end{figure}

In this Letter, we answer the above questions by constructing a new quantization rule that directly connects Landau levels with zero-field magnetic response functions (see Eq.(\ref{eq_QR})) and that is accurate to any polynomial order of magnetic field in principle. It is equivalent to an interesting intersection theorem: at zero temperature the smooth density of carriers as a function of Fermi energy from the semiclassical theory intersects the step-wise one from quantum mechanics almost at the middle point of each vertical jump (see Fig.(\ref{fig_int})). The deviation from those middle points is exponentially small at small magnetic field.

The implication of this quantization rule is two-fold. On one hand, the lowest order relationship naturally gives the fermi surface area as before, and corrections from Berry phase and magnetic moment are now summarized together in terms of the zero-field magnetization of the system.  Further corrections involve successively the magnetic susceptibility and even higher order magnetic response functions. Therefore, a knowledge of the magnetic response functions as calculated from semiclassical theories, allows accurate determination of the Landau levels.  On the other hand, measurement of the Landau levels from Shubnikov-de Haas oscillations, for example, can enable one to determine the zero-field magnetization, susceptibility and higher order magnetic response, which is otherwise hard to achieve for two dimensional systems.

\begin{figure}
\setlength{\abovecaptionskip}{0pt}
\setlength{\belowcaptionskip}{0pt}
\scalebox{0.22}{\includegraphics*{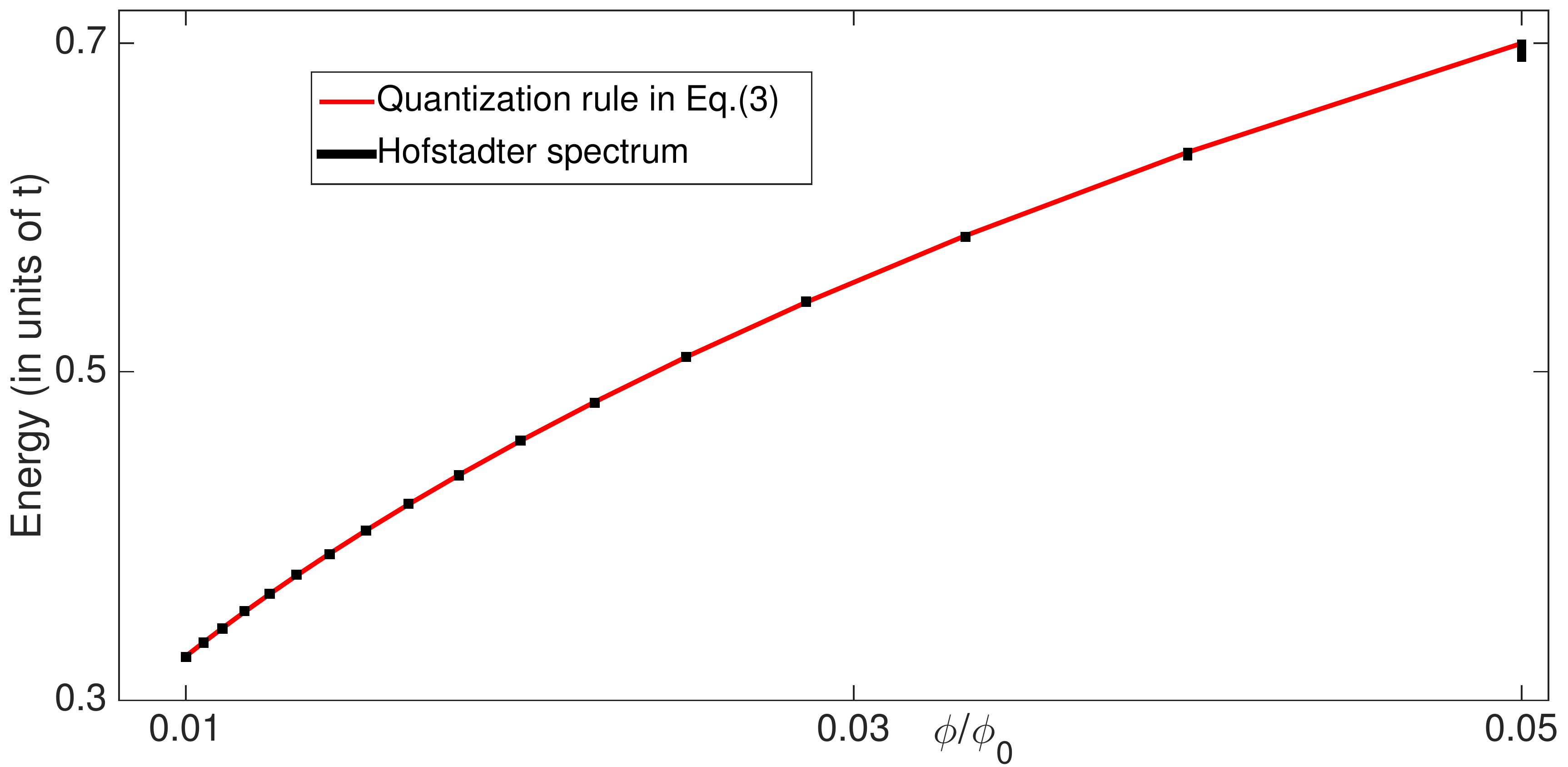}}
\caption{(color online) Hofstadter spectrum and the Landau levels based on Eq.(\ref{eq_QR}).}
\label{fig_hof}
\end{figure}

\begin{figure}[t]
\setlength{\abovecaptionskip}{0pt}
\setlength{\belowcaptionskip}{0pt}
\scalebox{0.22}{\includegraphics*{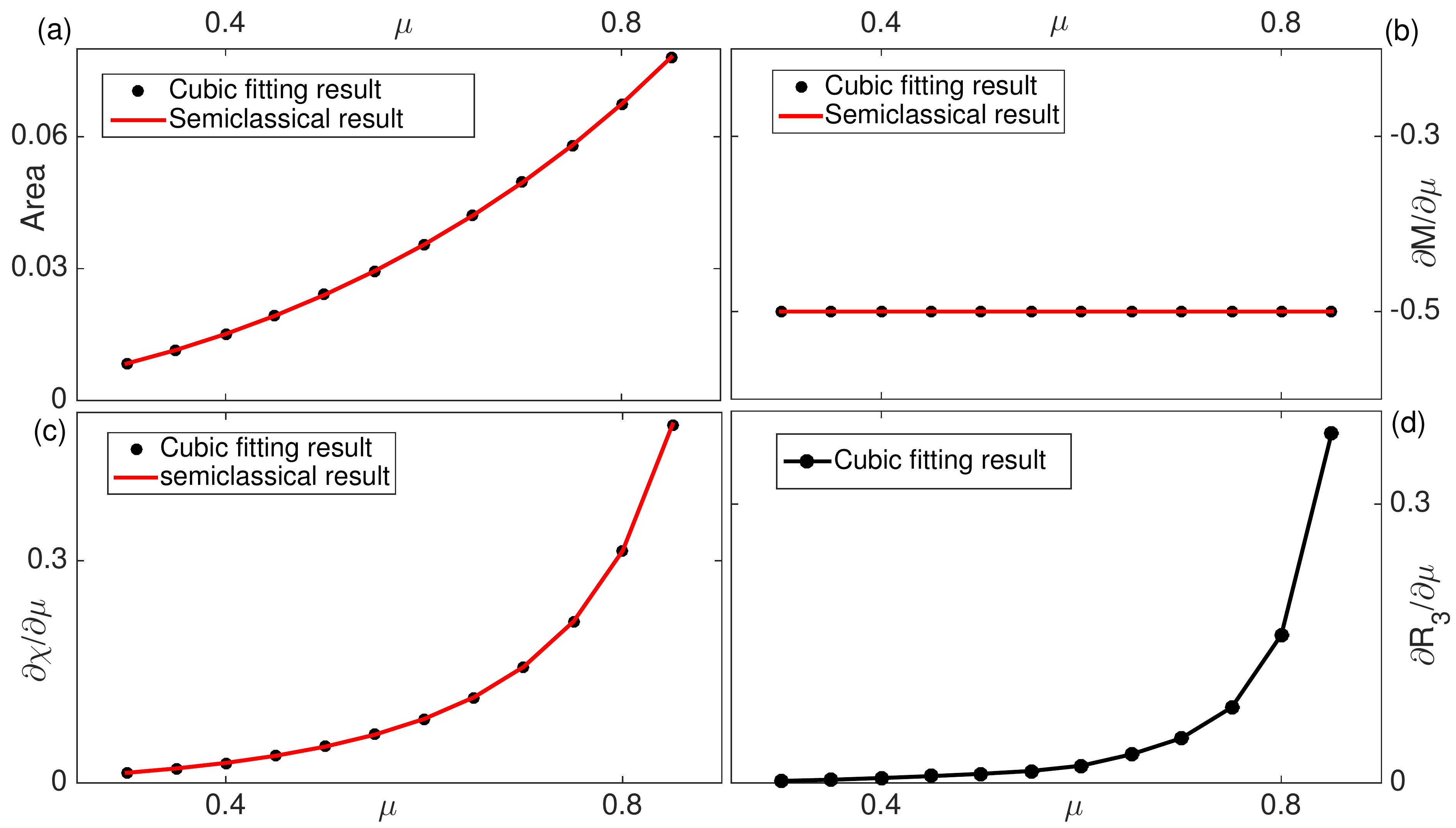}}
\caption{(color online) Obtain the equal-energy contour area and the derivatives of the magnetization, the susceptibility and the third order response $R_3$ from the Hofstadter spectrum. In Panel (a-d), y-axis are in units of the Brillouin zone area $4\pi^2/A$, $1/\phi_0$, $2A/\phi_0^2$, and $6A^2/\phi_0^3$, respectively. $\mu$ are in units of $t$.}
\label{fig_exp}
\end{figure}

As an example, in the low energy model of double layer graphene, by using the orbital magnetization and susceptibility,  our theory indeed produces Landau levels accurate to first and second order, respectively.  We also apply our theory to lattice models, and obtain Landau levels consistent with the Hofstadter spectrum (see Fig.(\ref{fig_hof})). Moreover, in Fig.(\ref{fig_exp}), we illustrate the inverse process of extracting the Fermi surface and the derivatives of the zero-field magnetization and susceptibility from the Hofstadter spectrum based on our quantization rule. They agree well with predictions of the semiclassical theory.

{\it Landau level quantization rule.}--- Without loss of generality we first consider Landau levels from the band minimum in two dimensions.
There is a heuristic picture of the quantization rule, and it starts from the fact that each Landau level carries the same density of states in two dimensional electron gas. When the Fermi energy falls inside the gap between neighboring Landau levels, the transverse conductivity is quantized (the quantum Hall effect) \cite{Klitzing1980}: $\sigma_{\rm trans}=ne^2/h$, where $n$ is an integer called Landau level filling factor. The transverse current from this conductivity can also be derived by multiplying the density of states $\rho_{\rm g}$ with the drift velocity: $\bm j=e\rho_{\rm g} (\bm E\times \bm B)/B^2$, where $\bm E$ is in-plane electric field and $\bm B$ is out-of-plane magnetic field. Therefore, the density of states is also quantized $\rho_{g}=neB/h$, and each Landau level has the same density of states $eB/h$.

Now we calculate the total density of states at the limit of zero temperature when the Fermi energy coincides with the $n$-th Landau level ($n$ starts from $0$).
Notice that the lower $n$ levels are beneath the Fermi energy, contributing a total carrier density $neB/h$, while the $n$-th level itself coincides with the Fermi energy and is considered as half occupied as temperature approaches zero due to the fact that the Fermi distribution function is exactly $1/2$ at the Fermi energy. Therefore, the total density of states reads ($\phi_0=h/e$ is the flux quanta):
\begin{equation}\label{eq_density1}
\rho_{\rm quan}=\left(n+{1\over 2}\right){B\over \phi_0}\,.
\end{equation}

We can also calculate the carrier density from the semiclassical theory through the statistical mechanics: 
\begin{equation}\label{eq_density2}
\rho_{\rm semi}=-\left. {\partial G\over \partial \mu} \right |_{B,T}\,
\end{equation}
where $G$ is the grand potential, T is the temperature, and $\mu$ is the chemical potential. Under a magnetic field, the semiclassical grand potential is always an analytic function of $B$ and can be expanded in terms of magnetic response functions: $G=G_0-BM-(1/2) \chi B^2+\cdots$, where $G_0$ is the grand potential at $B=0$, $M$ is the orbital magnetization, $\chi$ is the orbital magnetic susceptibility, and so on.

If we naively equate Eq.(\ref{eq_density1}) to Eq.(\ref{eq_density2}), we obtain the following quantization rule for Landau levels: at zero temperature when $\mu$ falls on the Landau level with an index $n=0,1,\cdots$,
\begin{equation}\label{eq_QR}
\left(n+{1\over 2}\right){B\over \phi_0}=-{\partial G_0\over \partial \mu}+B{\partial M\over \partial \mu}+{1\over 2} B^2 {\partial \chi\over \partial \mu}+\cdots \,.
\end{equation}
However, we emphasize that the density of states on the left hand side is valid when we take the limit of $T\rightarrow 0$ at a finite magnetic field, while the magnetic response functions on the right hand side is calculated by taking the limit $B\rightarrow 0$ and then $T\rightarrow 0$. We need to reconcile this contradiction to validate Eq.(\ref{eq_QR}).

The justification of Eq.(\ref{eq_QR}) is as follows: if we initially calculate those zero-field magnetic response functions from the semiclassical theory and solve a Landau level spectrum from Eq.(\ref{eq_QR}), we can construct a free energy of this spectrum from quantum mechanics. Then surprisingly, the resulting zero-field magnetic response functions are the same as those used initially. In other words, Eq.(\ref{eq_QR}) generates a Landau level spectrum that is consistent with magnetic response functions from the semiclassical theory. The proof is provided in \cite{suppl}.

Moreover, the Landau level spectrum from Eq.(\ref{eq_QR}) is quite accurate \cite{suppl}: if various magnetic response functions are given, then by construction the resulting Landau level spectrum cannot differ from the exact one by any polynomial of $B$ when $B$ is small; if we otherwise cut the right hand side of Eq.(\ref{eq_QR}) at some finite order of $B$, the resulting Landau level spectrum is guaranteed to be accurate to the same order of $B$.

Eq.(\ref{eq_QR}) can be reformulated to a delightful intersection theorem. At zero temperature, the total density of states as a function of energy is a step-wise function: it experiences a constant jump $B/\phi_0$ at each Landau level energy. The left hand side of Eq.(\ref{eq_QR}) finds the middle point of each vertical jump, while the right hand side yields a smooth density of states as a function of both magnetic field and Fermi energy. Therefore, Eq.(\ref{eq_QR}) indicates that this smooth semiclassical density of states always intersects the step-wise one near those middle points with an exponentially small deviation at small magnetic field. This is illustrated in Fig.(\ref{fig_int}). To our knowledge, this connection between these two densities of states is first recognized here.

Eq.(\ref{eq_QR}) can be easily generalized for Landau levels of holes. In this case, we have a quantization of the empty states. Therefore, the magnetic response functions on the right hand side of Eq.(\ref{eq_QR}) must be calculated from the band maximum, i.e. we need to replace the Fermi distribution function $f$ to $1-f$.

Eq.(\ref{eq_QR}) is also applicable to complicated systems which have multiple disconnected Fermi surfaces. Then we need to identify different sets of Fermi surfaces, so that Fermi surfaces in each set can shrink continuously to the same band maximum or minimum. To obtain corresponding Landau levels for each set of Fermi surfaces, we can apply Eq.(\ref{eq_QR}) with the magnetic response functions that contains the contribution only from this set of Fermi surfaces. This is illustrated for a lattice model later and for a spin-orbit coupling model in supplement \cite{suppl}.

{\it The Onsager's rule and beyond.}--- 
Eq.(\ref{eq_QR}) is the main result of our work and we demonstrate its significance through its relation with the Onsager's rule. First, we truncate its right hand side at the zeroth order term. Since  $-\partial G_0/\partial \mu=S_0/(4\pi^2)$ where $S_0$ is the $k$-space area enclosed by the equal-energy contour in the band structure $\varepsilon_0$ with the energy $\mu$, we obtain the Onsager's rule for Landau levels of electrons: $S_0=2\pi \left(n+{1\over 2}\right) {eB\over \hbar}$.

Then we truncate Eq.(\ref{eq_QR}) at the first order term $B(\partial M/\partial \mu)$. We consider the spinless case for simplicity and the orbital magnetization $M$ contains two contributions, one from the orbital magnetic moment $m$ and the other one from the Berry curvature $\Omega$ \cite{Xiao2005, Shi2007}: $M=\int (mf-\Omega g) d^2k/(4\pi^2)$. Here $f=dg/d\varepsilon$ is the Fermi distribution function. Therefore, 
\begin{equation}\label{eq_mag}
{\partial M\over\partial \mu}=-\int (mf^\prime-\Omega f) {d^2k\over 4\pi^2}\,. 
\end{equation}
If we combine the first term with $-\partial G_0/\partial \mu$ and move the second term to the left hand side of Eq.(\ref{eq_QR}), we obtain the following quantization condition: 
\begin{equation}\label{eq_qr_1st}
S^\prime =2\pi \left(n+{1\over 2}-{\Gamma(\mu)\over 2\pi}\right) {eB\over \hbar}\,,
\end{equation}
where $S^\prime=\int f(\varepsilon_0-Bm-\mu) d^2k$ is the area enclosed by the equal-energy contour in the modified band structure $\varepsilon_0-Bm$ with the energy $\mu$, and $\Gamma(\mu)$ is the Berry phase associated with the semiclassical orbit (note that the difference between Berry phases of equal-energy contours in $\varepsilon_0$ and $\varepsilon_0-Bm$ only contributes at second order in Eq.(\ref{eq_qr_1st})). This is exactly the modified Onsager's rule due to the Berry phase and magnetic moment effect \cite{Chang1996, Xiao2010, Wilkinson1984,Mikitik1999}. 

Even though the equivalence at first order, our quantization rule still has a few advantages over the modified Onsager's rule in previous work. First, it not only confirms that the Berry phase effect is not enough for the first order correction to the Onsager's rule, as illustrated for some specific models in previous work \cite{Carmier2008, Fuchs2010}, but also offers a concrete and general form where the orbital magnetization $M$ is the essential ingredient as a combination of the Berry phase and orbital magnetic moment effect.
Moreover, since response functions in Eq.(\ref{eq_QR}) only contain the Fermi function in the original band $\varepsilon_0$, the equal-energy contour in $\varepsilon_0$ is required. This is a computational advantage over directly finding the equal-energy contour of $\varepsilon_0-Bm$ as in Eq.(\ref{eq_qr_1st})\cite{Chang1996}, since $m$ can be dramatically enhanced around the point where two bands are near each other, creating significant numerical error in the equal-energy-contour finding process.

We comment that naively generalizing the Berry phase and magnetic moment to higher order in magnetic field does not guarantee a correct Landau level spectrum at higher order. Actually, if we truncate Eq.(\ref{eq_QR}) at the second order term, it is the susceptibility that contributes and it is different from this naive generalization \cite{Yang2015}.

It is interesting to point out that Eq.(\ref{eq_density2}) and its expansion indicate the following Maxwell relations: $(\partial \rho/\partial B)|_{B=0}=\partial M/\partial \mu$ and $(\partial ^2\rho/ \partial B^2)|_{B=0}=\partial \chi/\partial \mu$. Combined with Eq.(\ref{eq_mag}) and compared with the St\v{r}eda formula \cite{Streda1982}, the first Maxwell relation suggests that the anomalous Hall conductivity $\sigma_{xy}=\int \Omega f d^2k/(4\pi^2)$ is merely part of the contribution to $\partial \rho/\partial B$, and the orbital magnetic moment $m$ is also important.

In summary, Eq.(\ref{eq_QR}) interprets the Onsager's rule as from the zeroth order contribution to the free energy and modification from the Berry phase and magnetic moment as from the first order magnetic response function, i.e. the zero-field magnetization. It also suggests that further corrections to Onsager's rule corresponds to successively higher order magnetic response functions.

{\it Application in continuum and lattice models.}--- As a concrete example, we consider the continuum model in double layer graphene \cite{Mccann2006} (for simplicity, we choose $e$, $\hbar$ and the effective mass to be unity): 
$\hat{\rm H}=-{k_1^2-k_2^2\over 2} \sigma_1-{k_1 k_2} \sigma_2+\Delta \sigma_3$,
where $k_1$ and $k_2$ are momentum in $x$ and $y$ direction, $2\Delta$ is the band gap, and $\sigma_1$, $\sigma_2$ and $\sigma_3$ are Pauli matrices. Under a $B$ filed along the $z$ direction, this model can be solved exactly and Landau levels in the conduction band are: $\varepsilon_{\rm quan}=\sqrt{\Delta^2+n(n-1)B^2}$, with $n=0,1,2,\cdots$.

Now we apply our quantization rule in Eq.(\ref{eq_QR}) up to second order. The conduction band dispersion is $\varepsilon_{\rm c}=\sqrt{\Delta^2+k^4/4}$. Therefore, the equal-energy contour is a circle with the area $S_0=\pi k^2$. The magnetization and susceptibility can be easily calculated from the semiclassical theory \cite{Yang2015}, with $\partial M/\partial \mu=1/(2\pi)$ and $\partial \chi/\partial\mu=1/(8\pi \sqrt{\mu^2-\Delta^2})$. With the above quantities, Eq.(\ref{eq_QR}) yieds:
\begin{equation}\label{eq_quan_dlg}
\left(n+{1\over 2}\right) B= \sqrt{\varepsilon_2^2-\Delta^2} + {B\over 2}+{B^2\over 8} {1\over \sqrt{\varepsilon_2^2-\Delta^2}}\,.
\end{equation}
Solving the above equation by including the zeroth, first and second order terms on its right hand side and expanding the results at a small magnetic field, we obtain
$\varepsilon_0 =\varepsilon_{\rm quan}+B\sqrt{1-\Delta^2/\varepsilon_{\rm quan}^2}+O(B^2)$, 
$\varepsilon_1=\varepsilon_{\rm quan}+B^2/8+O(B^3)$, and
$\varepsilon_2=\varepsilon_{\rm quan}+O(B^3)$. These results clearly demonstrates that our quantization rule can correct the Onsager's rule order by order if corresponding magnetic response functions are known. In comparison, as shown in \cite{suppl}, the direct generalization of the Berry phase and magnetic moment to higher order cannot yield correct Landau levels up to second order.

Moreover, Eq.(\ref{eq_QR}) sometimes yields the exact Landau levels after truncated at a finite order. As an example, we consider the Dirac model in graphene: $\hat{H}=vk_1\sigma_1+vk_2\sigma_2+\Delta \sigma_3$, where $v$ is the Fermi velocity. Now we apply our quantization rule in Eq.(\ref{eq_QR}) up to the first order. The conduction band dispersion is $\varepsilon_{\rm c}=\sqrt{\Delta^2+v^2k^2}$. Since $\partial M/\partial \mu=1/(4\pi)$, Eq.(\ref{eq_QR}) yields the exact Landau levels $\varepsilon_{\rm quan}=\sqrt{\Delta^2+2nv^2B}$. This coincidence implies that all orbital magnetic response functions at higher than first order for this model must be a constant in $\mu$, which agrees with the exact quantum mechanical calculation \cite{suppl}.

An interesting comment is that in the above two models, $\partial M/\partial\mu$ are both constants, causing constant shifts of the level index: $n+1/2\rightarrow n-1/2$ in the low energy model in double layer graphene and $n+1/2\rightarrow n$ in the Dirac model in graphene. Our theory suggests that such constant shifts are dominated by the Berry phase effect for higher levels away from the band edge and by the orbital magnetic moment for lower levels near the band edge. The Berry phase effect contributes alone only when the band gap vanishes.

Our quantization rule can be easily applied to the lattice model as well. As an example, we consider the following tight-binding graphene model \cite{NG2009}: $\hat{H}=-t\sum_{\langle i,j\rangle}c_{i}^\dagger c_{j}$, where $t$ is the strength of the nearest neighbor hopping. We will focus on the conduction band. To obtain Landau levels, we analyze the topology of the equal-energy contour first. The conduction band has two degenerate valleys around two energy minimum with $\varepsilon=0$ at $K$ and $K^\prime$ point, one saddle point with $\varepsilon=t$, and one maximum with $\varepsilon=3t$ at the $\Gamma$ point. According to the Morse index theorem \cite{Bott1982}, 
there are two sets of degenerate electron-like semiclassical orbits at the range $0<\varepsilon<t$ which enclose $K$ and $K^\prime$ point separately, and one set of hole-like semiclassical orbits at the range $t<\varepsilon<3t$ which enclose the $\Gamma$ point. Different sets of semiclassical orbits yield Landau levels with different sets of level index.

With the above analysis in mind, we calculate magnetic response functions only from one valley from the semiclassical theory \cite{Yang2015}, and use Eq.(\ref{eq_QR}) up to second order to obtain the zeroth Landau level. In Fig.(\ref{fig_hof}), we compare the results with the Hofstadter spectrum at small magnetic field range, and they agree very well. 

We comment that mini-bands in Hofstadter spectrum has finite bandwidths. This corresponds to the fact that Landau levels are highly degenerate and tunneling between degenerate orbits broadens Landau levels \cite{Wilkinson1984}, which is beyond the scope of our theory. However, as discussed in previous works \cite{Kohn1958, Wilkinson1984, Guill1989}, the resulting bandwidth is only exponentially small at small magnetic field except near the saddle point where the semiclassical theory breaks down anyway. Therefore, this broadening effect does not introduce any contradiction to our theory.

{\it Extracting different magnetic responses.}--- As discussed above, our quantization rule shows that the Landau level formation is closely related to magnetic response functions. Therefore, a good knowledge of the Landau levels not only determines 
the fermi surface area, but also the magnetization, the susceptibility, etc. The key step is to obtain derivatives of response functions from the Landau level fan diagram or Hofstadter spectrum. Then their numerical integration over the chemical potential yields corresponding response functions. 

In the Landau level fan diagram, the method proceeds as follows: we first track discrete Landau levels at the same gate voltage, i.e. the same chemical potential, obtaining their level index $n$ and corresponding magnetic fields; then those data points are plotted in the diagram with the magnetic field $B$ as the $x$-axis and $(n+1/2)B$ as the $y$-axis, by fitting them to different orders of polynomials, we obtain derivatives of response functions through corresponding coefficients according to Eq.(\ref{eq_QR}).

The process is a little complicated for the Hofstadter spectrum which is discrete. We still focus on the range of spectrum with sufficiently small bandwidth. Usually if we fix an energy, we are not guaranteed to find a corresponding mini-band. However, as illustrated in Fig.(\ref{fig_hof}), the discrete spectrum can be well captured by the continuous semiclassical Landau levels. Therefore, for a fixed energy, we can obtain the corresponding flux through interpolations of the nearby spectrum data with the same band index. Then through the same process as discussed for the Landau level fan diagram, we can obtain derivatives of response functions.

As a concrete example, in Fig.(\ref{fig_exp}) by fitting the data from the pure case part in the Hofstadter spectrum of tight-binding graphene to cubic polynomials as described above, we obtain the Fermi surface area, $\partial M/\partial\mu$, $\partial \chi/\partial\mu$, and $\partial R_3/\partial \mu$ ($R_3=-\partial^3 G/\partial B^3$ is the third order response) from corresponding coefficients according to Eq.(\ref{eq_QR}). We find that the fitting results from the Hofstadter spectrum agrees well with the exact result directly from the semiclassical theory. We are also able to extract information about the third order response $R_3$ for the first time, and find that $R_3$ is nearly a constant at the band bottom but becomes large near the saddle point. Another interesting fact is that $\partial M/\partial \mu=-1/2$ (in units of $1/\phi_0$) identically, due to the fact that the orbital magnetic moment $m$ vanishes and only the identical Berry phase contributes. We emphasize that the above $\partial M/\partial \mu$ and $\partial \chi/\partial \mu$ are only contributions from one valley. We need to proceed similarly for the other valley to obtain the total contribution.

Finally, we comment that in experiments, impurities tend to localize current carrying states and hence broaden Landau levels. But the extended Landau levels reside in the band center. However, as pointed out previously, they can float up \cite{Laughlin1984, Koschny2003, Potempa2001} or go down \cite{Liu1996} in energy as $B$ is sufficiently small or the disorder is sufficiently strong, experiencing transitions from the quantum Hall state to the insulator state. Our quantization rule is validate only in the quantum Hall regime.

\begin{acknowledgments}
We acknowledge useful discussions with H. Chen,  J. Zhou, X. Li, R. Cheng, and L. Zhang. QN is supported by NBRPC (No. 2012CB921300 and No. 2013CB921900), and NSFC (No. 91121004). YG is supported by DOE (DE-FG03-02ER45958, Division of Materials Science and Engineering) and Welch Foundation (F-1255).
\end{acknowledgments}

\bibliographystyle{apsrev4-1}

\end{document}